\documentclass{mem}
\usepackage{natbib}\usepackage{txfonts}\usepackage{balance}
\usepackage{graphicx}
\usepackage[a4paper]{hyperref}
\idline{75}{282}
\begin{document}
\def\teff{$T\rm_{eff }$}
\def\kms{$\mathrm {km s}^{-1}$}

\newcommand{\COBOLD}{\mbox{CO$^5$BOLD}}
\newcommand{\Teff}{\mbox{$T_{\mathrm{eff}}$}}
\newcommand{\logg}{\mbox{$\log g$}}
\newcommand{\moh}{\ensuremath{\left[\mathrm{M/H}\right]}}
\newcommand{\Hpsurf}{\ensuremath{H^\mathrm{surf}_\mathrm{p}}}

\title{The CIFIST 3D model atmosphere grid}
\subtitle{}

\author{
H.-G.~Ludwig\inst{1,2}
\and
E.~Caffau\inst{2}
\and
M.~Steffen\inst{3} 
\and
B.~Freytag\inst{1,2,4} 
\and
P.~Bonifacio\inst{1,2,5}
\and
A.~Ku\v{c}inskas\inst{6,7}
}

\offprints{H.-G. Ludwig}
 
\institute{
CIFIST -- Marie Curie Excellence Team
\and
GEPI -- Observatoire de Paris,  CNRS, Universit{\'e} Paris Diderot, 92195 Meudon, France
\and
Astrophysikalisches Institut Potsdam, An der Sternwarte 16, 14482 Potsdam, German
\and
CRAL --  UMR 5574 CNRS, Universit{\'e} de Lyon, {\'E}cole Normale Sup{\'e}rieure de Lyon,
46~all{\'e}e d'Italie, 69364~Lyon Cedex~07, France
\and
INAF -- Osservatorio Astronomico di Trieste, via Tiepolo 11, 34143 Trieste, Italy
\and
Institute of Theoretical Physics and Astronomy, Go\v{s}tauto 12, Vilnius
LT-01108, Lithuania 
\and
Vilnius University Astronomical Observatory, \v{C}iurlionio 29, Vilnius LT-03100, Lithuania
}

\authorrunning{Ludwig et al. }

\titlerunning{The CIFIST 3D model atmosphere grid}

\abstract{%
  Grids of stellar atmosphere models and associated synthetic spectra are
  numerical products which have a large impact in astronomy due to their
  ubiquitous application in the interpretation of radiation from individual
  stars and stellar populations. 3D model atmospheres are now on the verge of
  becoming generally available for a wide range of stellar atmospheric
  parameters. We report on efforts to develop a grid of 3D model atmospheres
  for late-type stars within the CIFIST Team at Paris Observatory. The
  substantial demands in computational and human labor for the model
  production and post-processing render this apparently mundane task a
  challenging logistic exercise. At the moment the CIFIST grid comprises 77 3D
  model atmospheres with emphasis on dwarfs of solar and sub-solar
  metallicities. While the model production is still ongoing, first applications
  are already worked upon by the CIFIST Team and collaborators.  
\keywords{Stars: abundances -- Stars:
    atmospheres -- hydrodynamics -- convection -- radiative transfer} }
\maketitle{}

\section{The simulation code} 

The 3D simulations were performed with the
radiation-hydrodynamics code \COBOLD\ \citep{Freytag+al02,Wedemeyer+al04}.
The code solves the time-dependent equations
of compressible hydrodynamics coupled to radiative transfer in a constant
gravity field in a Cartesian computational domain which is representative of a
volume located at the stellar surface.
The equation of state takes into consideration the ionization of hydrogen
and helium, as well as the formation of H$_2$ molecules according to
Saha-Boltzmann statistics. Relevant thermodynamic quantities -- in
particular gas pressure and temperature -- are tabulated as a function of
gas density and internal energy.
The multi-group opacities used by \COBOLD\ are based on monochromatic opacities
stemming from the MARCS stellar atmosphere package \citep{Gustafsson+al08}
provided as function of gas pressure and temperature with high wavelength
resolution. For the calculation of the opacities solar elemental abundances
according \citet{Grevesse+Sauval98} are assumed
with the exception of CNO for which values
close to the recommendation of \citet{Asplund05} are adopted
(specifically, A(C)=8.41, A(N)=7.80, A(O)=8.67).

\begin{figure*}[t!]
\resizebox{0.48\hsize}{!}{\includegraphics[clip=true]{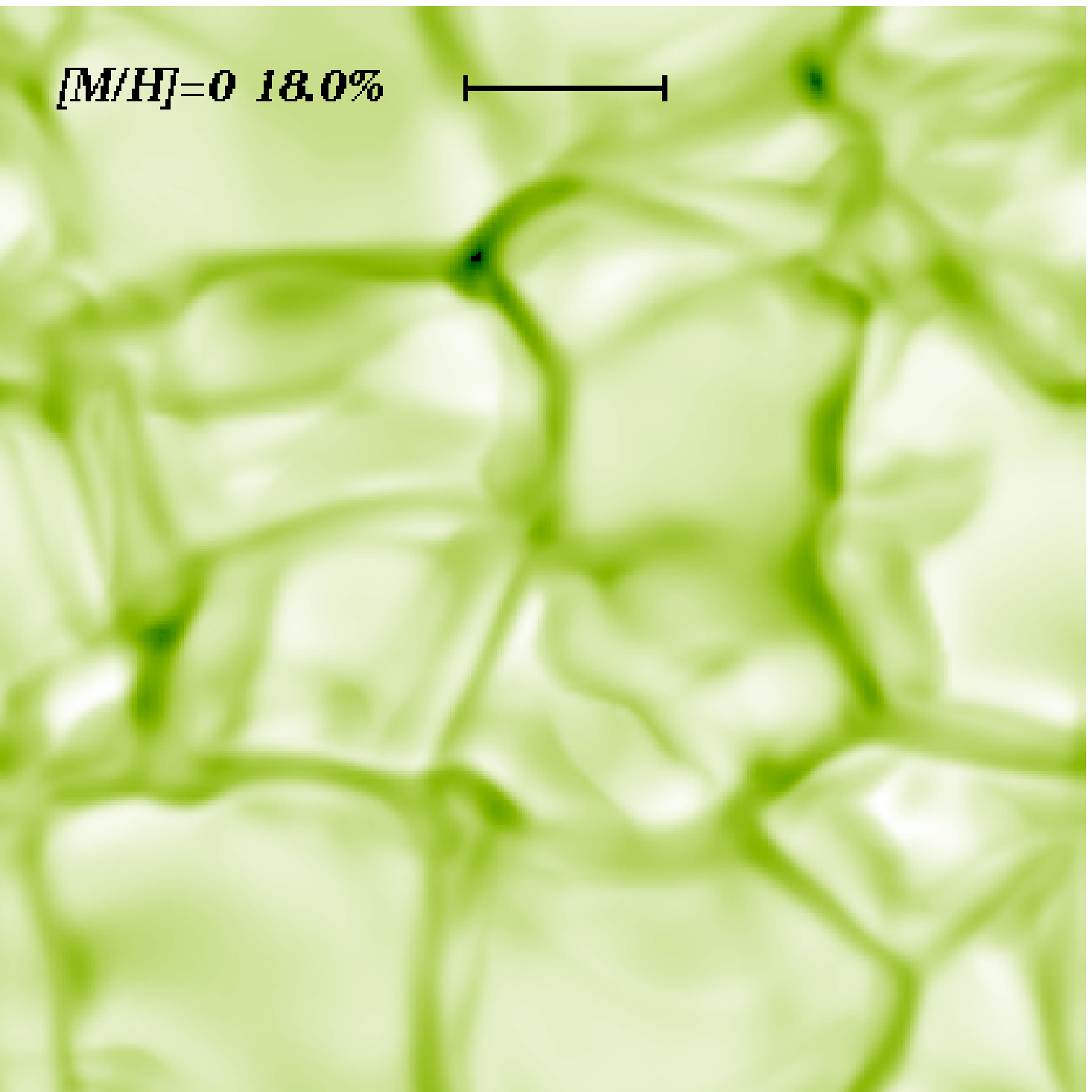}}\hfill
\resizebox{0.48\hsize}{!}{\includegraphics[clip=true]{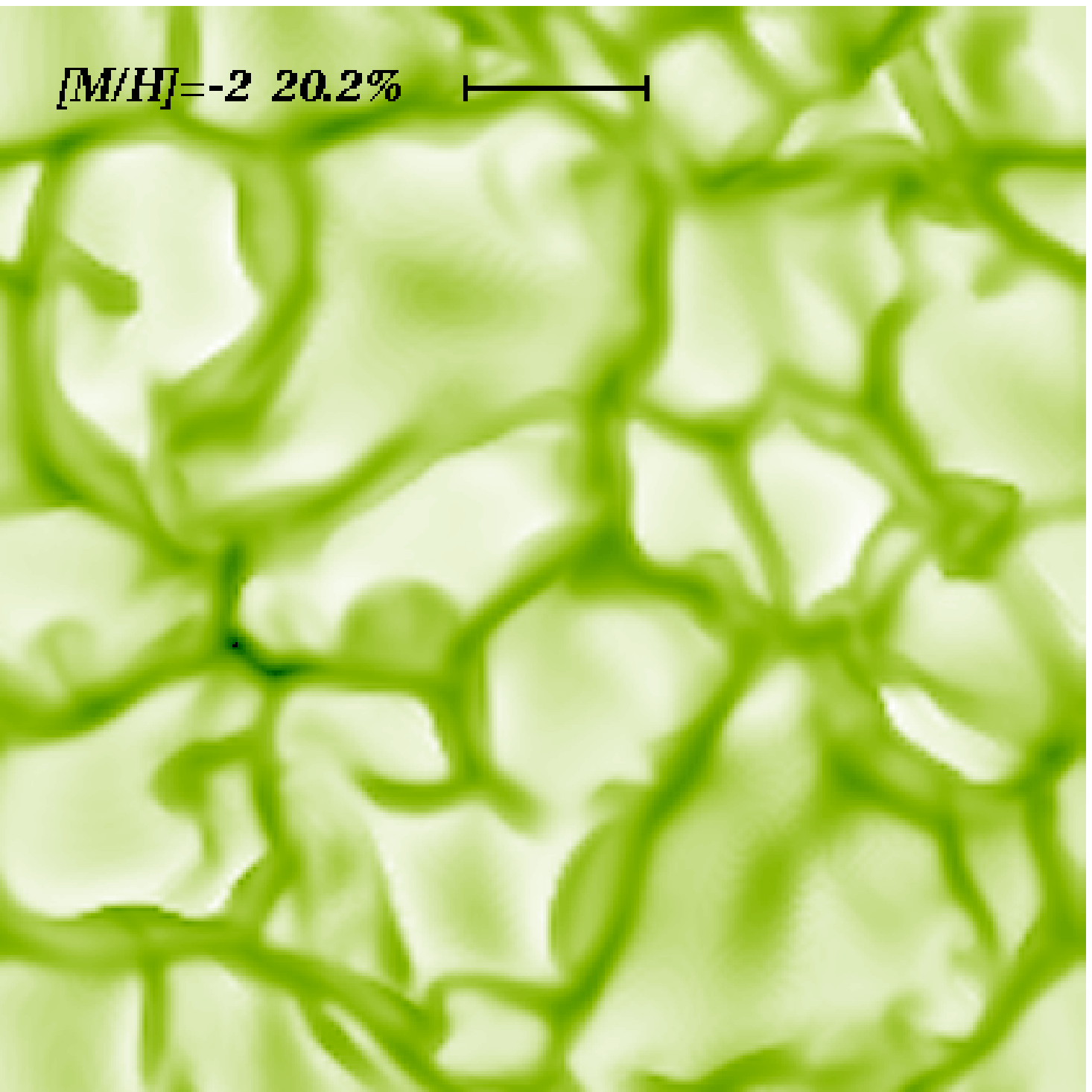}}
\caption{\footnotesize Typical snapshots of the emergent bolometric intensity
  (false color representation) from two runs at \Teff=5000\,K, \logg=2.5 and
  metallicties \moh=0 and -2. The metallicity, the relative RMS intensity
  contrast of the image and a scale of 10 times the pressure scale height at
  $\tau=1$ are given in each frame.}
\label{f:snaps}
\end{figure*}

\section{Model set-up} 

We typically used a number of $140\times 140 \times 150$ points for the
hydrodynamical grid. The wavelength dependence of the radiation field was
represented by 5 multi-group bins in the case of solar metallicity, and 6 bins
at sub-solar metallicities following the procedures laid out by
\citet{Nordlund82,Ludwig92,Ludwig+al94,Voegler+al04}. For test purposes we
calculated a few models using 12 bins. The sorting into wavelength groups was
done applying thresholds in logarithmic Rosseland optical depth $\{ +\infty,
0.0, -1.5, $ $-3.0, -4.5, -\infty\}$ for the 5-bin, and $\{+\infty, 0.1, 0.0,$
$-1.0, -2.0, -3.0, -\infty\}$ for the 6-bin schemes. In the case of 12 bins we
used as thresholds $\{ +\infty, 0.15, 0.0, $ $ -0.75, -1.5, -2.25, $ $-3.0,
-3.75, -4.5, -\infty\}$; in addition, the first three continuum-like bins were
split each into 2 bins according to wavelength at 550, 600, and 650\,nm. In
all but one bin a switching between Rosseland and Planck averages was
performed at a band-averaged Rossland optical depth of 0.35; in the bin
gathering the largest line opacities the Rosseland mean opacity was used
throughout. The decisions about number of bins, and sorting thresholds are
motivated by comparing radiative fluxes and heating rates obtained by the
binned opacities in comparison to the case of high wavelength resolution.

We constructed 3D starting models by scaling existing 3D models according to
1D standard stellar structure models. The 1D models provided scaling factors
for thermodynamic, kinematic, and geometrical properties to modified \Teff,
\logg, and metallicity. Points of the same optical depth and same pressure
relative to the surface pressure were considered associated between the
original and scaled model, in the optically thin and thick regions,
respectively. Empirically, the scaling procedure works particularly reliably
if the scaling is performed along lines of constant entropy jump in the
\Teff-\logg-plane \citep[see, e.g.,][]{Ludwig+al99}.

At given \Teff\ and \logg\ we kept the computational domain the same for each
metallicity. Since the models were intended to serve as atmosphere models for
spectral synthesis calculations, a minimum extent of the computational domain
to $\log\tau_\mathrm{Ross}=-6$ was always required.  The overall philosophy
was to keep the model set-up -- in a relative sense -- solar-like.  We
required a minimum time interval over which the model could be considered as
relaxed which corresponded to an equivalent of one hour in a solar model.

\section{Model calculations and monitoring} 

The majority of the simulation runs was conducted on a dedicated Linux cluster
of 14 double-processor, double-core nodes running one model per node making
use of OpenMP parallelization. A main-sequence run is completed in 1--2 months
(wall clock time). More CPU demanding giant models were calculated on an
IBM~SP5 at the Italian CINECA supercomputing center, and in addition on
machines at the Institute of Theoretical Physics and Astronomy of Vilnius
University (Lithuania). Running up to 20 hydrodynamical models in parallel
made it necessary to facilitate the monitoring of simulation runs.  For this
purpose we developed an automatic procedure producing diagnostic plots in
HTML-format depicting the evolution of the thermal, flux, granulation pattern,
and oscillatory properties of a model.  Figure~\ref{f:snaps} shows examples of
granulation patterns at different metallicity.

\section{Snapshot selection and auxiliary data.} 

After completion of a model run a sub-set of snapshots is selected for
spectral synthesis purposes. The selection is guided by the requirement that
the statistics of the sub-sample should closely resemble the statistics of the
whole run. In particular, the statistics of fluctuation in velocity and
emergent flux should be preserved. The selection is done for reducing the
amount of necessary calculations since 3D spectral synthesis calculations are
computationally demanding. The selected data is augmented by basic information
about mean model properties, and 1D standard atmosphere models which exactly
share the atmospheric parameters with the 3D run.

\section{Grid completeness and computational costs} 

\begin{figure*}[t!]
\begin{center}
\resizebox{0.79\hsize}{!}{\includegraphics[clip=true,angle=90]{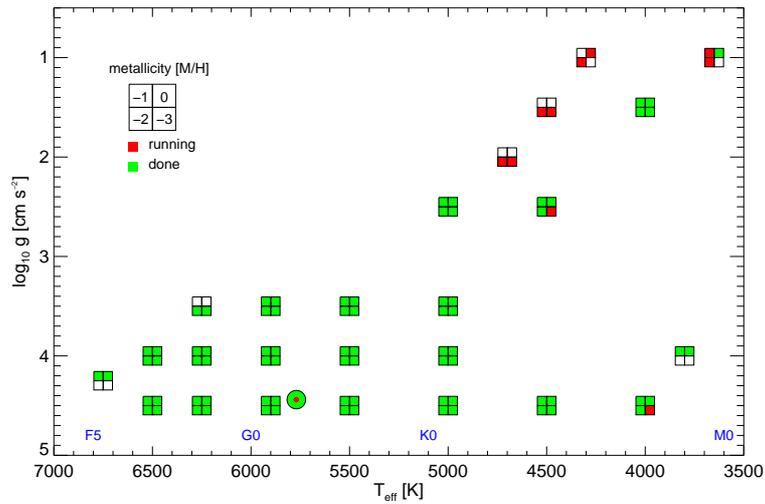}}
\caption{\footnotesize Status of the model production for the CIFIST grid.
  Symbols mark the location of a model in the \Teff-\logg-plane. Green color
  indicates completed model runs, red ongoing calculations. Each square is
  split into four sub-squares indicating solar, 1/10, 1/100, and 1/1000 of
  solar metallicity. The solar position is indicated by the round symbol.}
\end{center}
\label{f:grid}
\end{figure*}

\begin{figure}[]
\resizebox{\hsize}{!}{\includegraphics[clip=true,angle=90]{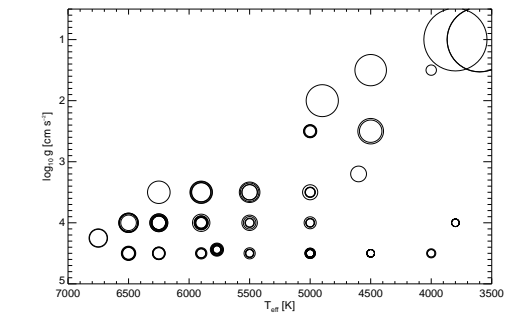}}
\caption{\footnotesize
Approximate computational model cost.
The surface area of a circles depicts the approximate compute time of a 
model run. For details see text. 
}
\label{f:costs}
\end{figure}

At the moment 77 non-solar 3D models have been completed (see
Fig.~\ref{f:grid}). On top of this number we calculated a series of solar
models and test models with more detailed (12 bins) radiative transfer. Due to
a combination of specific needs for projects and computational costs most
models were produced for dwarfs and sub-giants. Figure~\ref{f:costs}
illustrates the approximate computational costs -- i.e., the demand in CPU
time -- of a model run; the symbol diameter is proportional to the quantity
$\left(\Delta t/\Delta t_\mathrm{CFL}\right)^{-0.5}$ where $\Delta t$ is the
typical timestep of a model run, and $\Delta t_\mathrm{CFL}$ the
Courant-Friedrichs-Levy time.  Metallicity generally plays no dominant role
for the computational cost. The computational cost rises significantly towards
the red giant branch, and towards higher effective temperatures. This
basically reflects the decrease of the radiative time scale relative to
$\Delta t_\mathrm{CFL}$ leading to a reduction of the overall time step.

\section{Thoughts for a next generation 3D grid} 

At present the realism of 3D model atmospheres is limited by the
approximations necessary to keep the radiative transfer tractable.
Consequently, in next generation grids much effort will be put in refining the
radiative transfer. This will be accompanied by a -- likely modest -- increase
in spatial resolution. A perhaps less obvious property should be a higher
degree of homogeneity among the models. A global strategy for designing the
computational grid is desirable which takes into consideration the physical
structure and constraints given by the numerics.  This also relates to the way
opacities are grouped. A global strategy ensures an increased differential
accuracy of properties among models (e.g., granulation contrast, granular
scale, equivalent mixing-length parameter, equivalent width of synthetic
lines, etc.). This process will benefit from the already existing 3D models.
If grid completeness is important the computational cost of giants
necessitates to include them in the production early. Last but not least
dedicated efforts are necessary to make the computational results accessible
to the broader astronomical community. Extensive tables of synthetic colors
and spectra are the obvious data products which need to be delivered.

\begin{acknowledgements}
BF, EC, HGL, and PB acknowledge support from
EU contract MEXT-CT-2004-014265 (CIFIST).
\end{acknowledgements}

\bibliographystyle{aa}

\end{document}